\documentclass[11pt,preprint]{aastex}

\begin{document}

\title{Rating Growth of Scientific Knowledge and Risk from Theory
Bubbles}

\author{Abraham Loeb\\Institute for Theory \& Computation\\
Harvard University\\60 Garden St., Cambridge, MA 02138}

\begin{abstract}

In physics the value of a theory is measured by its agreement with
experimental data.  But how should the physics community gauge the
value of an emerging theory that has not been tested experimentally as
of yet?  With no reality check, a hypothesis like {\it string theory}
may linger for a while before physicists will know its actual value in
describing nature.

In this short article, I advocate the need for a website operated by
graduate students that will use various measures of publicly available
data (such as the growth rate of newly funded experiments, research
grants, publications, and faculty jobs) to gauge the future dividends
of various research frontiers. The analysis can benefit from past
experience (e.g. in research areas that suffered from limited
experimental data over long periods of time) and aim to alert the
community of the risk from future theory bubbles.

\end{abstract}

\bigskip
\bigskip
\bigskip

{\it ~\\``All great ideas are dangerous.''}

\noindent
{\it Oscar Wilde, De Profundis}

\section{Theory Bubbles}

The financial world shares a common theme with research in physics:
the values of its assets should ultimately reflect hard facts, but
because risks are inherent in any creative human activity, the future
value of evolving assets is subject to uncertainty and speculation.
At first sight, one may question the need for forecasting the future
of scientific research. Perhaps we should apply the rule of natural
selection to scientific theories and let the fittest theory survive on
its own.  However, the benefits of a global overview are obvious.
Similarly to the business world, a balanced assessment of the level
of risk and potential dividends for investing research time in
emerging research frontiers can increase the efficiency of the work
force, leading to stronger growth.

The investment of research time in strong intellectual assets is of
critical importance for beginning graduate students who wish to
establish their careers on a good foundation. Young researchers are
unaware of the full menu of optional research areas and the history
accompanying each of them. In the current research landscape, students
often have to rely on a word of mouth from their PhD advisor or
colleagues.

An illustrative example for a research field with evolving
intellectual assets is the study of the Cosmic Microwave Background
(CMB) anisotropies. The field started with theoretical work in the
1960s and exhibited gradual progress in experimental capabilities over
a period of several decades until experiments reached the sensitivity
threshold for a detection. As soon as the COBE satellite reported
conclusive evidence for the CMB anisotropies in 1992, subsequent
experimental work generated huge dividends for young researchers
(theorists and observers alike) who joined this field at that
time. But there are research frontiers on the opposite side of the
spectrum. With no reality check for three decades, a hypothesis like
{\it string theory} may linger for a while before physicists will know
its actual value in describing nature.  The existence of such an
unusual branch within the tree of physical sciences is a testimony to
the fact that fundamental physics has reached a mature state in which
most of the low-hanging fruit was picked up already and the elevated
fruit takes longer and longer to reach.

In the early phase of an emerging research field, when there are
limited experimental facts to test the validity of its underlying
theoretical ideas, the physics community needs a credit rating agency
similar to {\it Standard \& Poor's (S\&P)}, {\it Moody's Investor
Service} or {\it Fitch Ratings} in the financial world, that would
evaluate the future promise of the field.  Naively, who would be
better suited for rating the future promise of research frontiers than
the most senior physicists? The problem is that many of these
physicists are already invested in speculative frontiers whose promise
needs to be evaluated. This leads to a conflict of interests and
wishful thinking. The associated bias is reminiscent of the AAA rating
given by the financial rating agencies to securities from which they
benefited (as the agencies were paid by the companies who issued these
securities), and which ended up collapsing during the financial crisis
of 2007-9. In the physics world, a long-lived bias of this type could
lead to similarly devastating consequences, such as an extended period
of stagnation characterized by a large community of talented
physicists investing their research time in intellectual assets whose
actual value in terms of describing nature may be low.  These
symptoms define a {\it theory bubble}.

\section{The Need for a Credit Rating Website}

In order to help young researchers choose a research topic with
realistic expectations, it would be helpful to establish a website
which would use quantitative measures of publicly available data to
gauge the promise and future dividends of different research
frontiers.  The evaluation metric should factor in, with proper
weights, all the ingredients that ultimately make physics research
successful. Among these ingredients are: the existence of an
underlying self-contained theory from first principles, the potential
for experimental tests of this theory, and the track record of related
research programs.  The evaluation metric has to be pre-determined and
anchored in numbers.  Of course, factors like intellectual excitement
cannot be quantified, but as long as funding agencies are doing their
job and the integrity of the researchers can be trusted, the data
about the growth of a field should echo this excitement factor (elbeit
with a time delay).

The relevant data includes the level of funding allocated to
experiments ${F}_{\rm exp}$ and research grants ${F}_{\rm grants}$, as
well as the number of publications ${N}_{\rm pubs}$ and faculty jobs
${N}_{\rm jobs}$ within the particular research field of interest.
Another important ingredient is the quality of the underlying
theoretical framework ${T}$, which can be normalized to have a value
between $0$ and $1$, with $1$ representing a unique, self-contained
theory derived from first principles and $0$ representing pure
phenomenology with no theoretical understanding.  The simplest model
relates the rate of change in these variables to a linear combination
of their values.  For example, the publication rate is expected to
scale as a linear combination of the number of faculty jobs and the
available research funds.  Similarly, the quality of the theory would
improve at a rate that is a linear combination of the numbers of
experiments and faculty jobs, and the growth rate of jobs might be
proportional to the level of funding in the field.  In the simplest
linear model, one may combine the above variables into a vector,
$\vec{\bf v}=(T,F_{\rm exp}, F_{\rm grants},N_{\rm pubs},N_{\rm
jobs})$, whose growth rate is,
\begin{equation}
{d\over dt} \vec{\bf v}=\vec{\vec{\cal M}}\vec{\bf v} ,
\label{eq:rate}
\end{equation}
where $\vec{\vec{\cal M}}$ is a $5\times 5$ matrix with 25 constant
coefficients that need to be calibrated based on historical data on
the research frontier of interest or similar ones. The eigenvectors of
the matrix $\vec{\vec{\cal M}}$ satisfy $d\vec{\bf v}/dt \propto
\vec{\bf v}$ and therefore grow exponentially in time with a growth
rate equal to their eigenvalue. One could rank these eigenvectors in
order of decreasing eigenvalues, with the top eigenvector indicating
the optimal mix of theory, experimental work and grants that the
research area needs in order to acheive its fastest exponential
growth. If this mix happens to be common among many different
sub-fields, then it can be used by funding agencies to decide whether
any particular sub-field does not match the proper universal mix
(e.g., by having a theory level that is too weak for its experimental
effort) and needs nurturing through a revised funding scheme that
would compensate for its weaknesses, similarly to a baby whose diet
needs additional nutrients in order for its body to be healthy and
grow at an ideal rate.

More complicated growth algorithms with nonlinear (e.g. power-law)
scalings may also be devised, although they often require a larger number
of free parameters that need to be calibrated. An example for a
nonlinear algorithm is,
\begin{equation}
{d\over dt} \vec{\bf v}=\vec{\vec{\cal M}}\vec{\bf p} ,
\label{eq:power}
\end{equation}
where the power-law vector, $\vec{\bf p}\equiv (T^\alpha,F_{\rm
exp}^\beta, F_{\rm grants}^\gamma,N_{\rm pubs}^\delta,N_{\rm
jobs}^\epsilon)$, adds five new constants that need to be calibrated:
$\alpha,\beta,\gamma,\delta$ and $\epsilon$. 

The migration of researchers or resources from one sub-field to
another can serve as an additional indicator for the future evolution
of $\vec{\bf v}$ in both sub-fields.  The algorithm can therefore be
expanded to incorporate the correlations among different sub-fields by
writing the right-hand-side of equations (\ref{eq:rate}) and
(\ref{eq:power}) as a sum of similar terms over different research
areas.

An alternative, parameter-free algorithm relies on measuring $\vec{\bf v}$
at three (or more) different times and extrapolating its value to the
future based on a Taylor expansion in time derivatives up to second
(or higher) order,
\begin{equation}
\vec{\bf v}(t) \approx \vec{\bf v}(0) + {\dot{\vec{\bf v}}}(0)t + {1\over
2}{\ddot{\vec{\bf v}}}(0)t^2 + ...
\end{equation}
where $\ddot{\vec{\bf v}}\equiv (d^2{\vec{\bf v}}/dt^2)$.  The
relative success of different algorithms can be assessed based on
their track record in predicting the evolution of various research
frontiers in historical data sets. This can be measured by comparing
the extrapolated values of $\vec{\bf v}(t)$ at late times $t$ to the
actual values of $\vec{\bf v}$ that are realized at those times. The
optimized algorithm serves as a collective learning tool, with the
potential of saving the physics community from the risk of repeating
unwanted circumstances.

Any individual researcher could project $\vec{\bf v}$ along some
particular unit vector of interest $\vec{\bf I}$ and measure the
success of any field through the scalars $S=\vec{\bf v}\cdot\vec{\bf
I}$ and $\dot{S}=\dot{\vec{\bf v}}\cdot\vec{\bf I}$. For a researcher
with a theoretical inclination, the projection vector $\vec{\bf I}$
might give more weight to ${T}$ than to ${F}_{\rm exp}$. On the other
hand, a researcher who is mostly concerned about employment
opportunities would put the largest weight on $\dot{N}_{\rm jobs}$.
An individual researcher may also prefer to normalize the value of $S$
by the number of researchers in a field and focus attention on
$s=(S/N_{\rm jobs})$, thus weakening the advantage of highly populated
fields where the fractional influence of an individual researcher
might be small. In motivating this normalization, one can argue that
the potential impact of an emerging field on the career of a beginning
physicist is inversely proportional to the (time-dependent) number of
competing physicists who are already working in this field. If many
physicists are engaged in the same project, as in experimental
particle physics (e.g. the {\it Large Hadron Collider}), it would be
more difficult for an entering physicist to leave a unique mark on the
overall progress of the project. Interestingly, $s$ may exhibit a
temporal decline even if all the elements of the matrix
$\vec{\vec{\cal M}}$ are positive definite and $S$ is monotonically
increasing.

Since the data used to calibrate the parameters of each algorithm have
error bars, the predictions inherit a range of uncertainty.  One can
imagine designing a ``stress test'' for the growth of a research field
allowing for one of the parameters to change drastically within its
range of uncertainty. Fields that depend on funding for a single
expensive experiment might be more vulnerable to dire outcomes, of the
nature that inflicted particle physics when the Superconducting
Super-Collider (SSC) had been canceled.

The growth rate of all research frontiers is affected by the global
research budget, which in turn is shaped by economic, industrial or
defense-related forces.  The global changes can be factored out by
focusing on the fractional differences between the values of $S$ in
different research areas.  The analysis in more advanced models could
include other external factors, such as technological advances which
lower the cost of experiments or computers.

The data required to calibrate the free parameters of the ranking
algorithms can be gathered through automated searches for keywords in
electronic data archives (such as {\it arXiv.org} or {\it NSF.org}).
The website may be updated once or twice a year. Aside from automated
searches, practitioners of fields that are being evaluated can submit
supplementary data that will be incorporated into the
analysis. Obviously, the next step in advancing this initiative would
be to use historical data in calibrating the above algorithms to best
match the evolution of particular research frontiers.

It would be most fitting for the rating website to be operated by
graduate students, since they would be the main consumers of its
recommendations and they also possess the lowest level of unwarranted
bias or prejudice.  The evaluation algorithms can benefit particularly
from past experience in research areas that suffered from limited
experimental data over extended periods of time. The main purpose of
the website is to extrapolate existing trends in research to the
future and alert the community to the risk of future theory
bubbles. Since surprises are inherent to scientific exploration and
innovative ideas are often under-funded, any predictive algorithm
would occasionally fail in a particular field, but that should not
take away the value of this endeavor in offering a global perspective
on the current state of mind in many other fields.

The existence of a balanced rating algorithm can also aid funding
agencies (such as NSF, DOE, or NASA) in optimizing their allocation of
funds to promote progress in physics research. These agencies would
naturally favor a rating algorithm that maximizes global scientific
returns rather than the benefits to the careers of individual
researchers. In fact, it would be in the interest of these funding
agencies to support the proposed website and provide incentives for
talented students (e.g. through special grants or fellowships) to keep
the statistical analysis which is featured in it at the highest
quality level.

Although graduate students may vote with their feet, senior physicists
bear the main responsibility for defining the directions of future
research. These senior researchers may artificially boost the values
of $F_{\rm grants}, N_{\rm pubs}$ and $N_{\rm jobs}$, but they cannot
honestly manufacture targeted experiments deserving a high $F_{\rm
exp}$ if those are not feasible and they cannot honestly claim to have
high values of $T$ if the theory is not understood.  Past experience
may therefore suggest putting the largest weights on $F_{\rm exp}$ and
$T$ in order to obtain a reliable forecast for the long-term evolution
of $\vec{\bf v}$.  The proposed website will be most effective if it
will convince senior researchers to shift their focus to new research
areas.  Such a shift could be naturally accomplished if the funding
agencies will be influenced by the rating procedure.  A strong
feedback could lead to an exponential growth of successful
disciplines. It is extremely important, however, that the funding
agencies will maintain balance and diversity among sub-fields, take
some risks, and avoid funneling most of the resources to a small
number of successful but conservative programs.\footnote{This point
was explained in more detail in Loeb, A., Nature {\bf 467}, 358
(2010); {\it http://arxiv.org/abs/1008.1586}.}

The analogy with the financial world has its limitations.  The stock
market feeds more vigorously on rumors, whereas scientific research
has a strong spine that is not as easily swayed by unsubstantiated
claims. Consequently, temporal changes in research trends are more
moderate.

\section{Concluding Remarks}

Mathematics is different from physics in that the values of its
intellectual assets are not measured by their match to experimental
facts.  Abstract aspects of theoretical physics could temporarily
evade experimental verification and masquerade as known truths, but
they could also go out of fashion as soon as new data rules them
out. The healthy interplay between abstract mathematical ideas and
data is essential for making progress in our never ending quest to
understand the one way in which nature is realized out of many
possibilities that it could have been.

\bigskip
\bigskip
\bigskip

\acknowledgements I thank Adrian Liu, Ido Liviatan, Oded Liviatan,
Tony Pan, Jonathan Pritchard, and Nick Stone for their helpful
comments on the manuscript.

\end{document}